\documentclass[letterpaper]{JHEP3}

\usepackage{epsfig}
\usepackage{amsmath}
\usepackage{xspace}

\def\be{\begin{equation}}
\def\ee{\end{equation}}
\def\beq{\begin{eqnarray}}
\def\eeq{\end{eqnarray}}
\def\a{\alpha}
\def\b{\beta}

\def\l{\lambda}

\def\({\left (}
\def\){\right )}
\def\[{\left [}
\def\[{\right ]}

\title{Violation of Energy Bounds in Designer Gravity}

\author{Thomas Hertog \\
Theory Division, CERN, CH-1211 Geneva 23, Switzerland \\
{\it and}\\
APC, 11 Place Marcelin Berthelot, 75005 Paris, 
France\thanks{UMR 7164 (CNRS, Universit\'e Paris 7, CEA, Observatoire de Paris)}\\
E-mail: \email{Thomas.Hertog@cern.ch}}

\date{July 2006}

\abstract{%

We continue our study of the stability of designer gravity theories, where one considers anti-de Sitter gravity 
coupled to certain tachyonic scalars with boundary conditions defined by a smooth function $W$. It has recently been 
argued there is a lower bound on the conserved energy in terms of the global minimum of $W$, 
if the scalar potential arises from a superpotential $P$ and the scalar reaches an extremum of $P$ at infinity. 
We show, however, there are superpotentials for which these bounds do not hold.

}

\preprint{\hepth{0607171}\\ CERN-PH-TH/2006-142} 

\begin{document}

%
\section{Introduction}
%

It is well known that anti-de Sitter (AdS) gravity coupled to a scalar with mass at or slightly above the 
Breitenlohner-Freedman (BF) bound \cite{Breitenlohner82} admits a large class of boundary conditions, defined by an 
essentially arbitrary real function $W$, for which the conserved charges are well defined and finite 
\cite{Henneaux02,Henneaux04,Hertog04,Hertog05,Henneaux06,Amsel06}. 

Theories of this type have been called designer gravity theories \cite{Hertog05}, because their dynamical properties depend 
significantly on the choice of $W$. For example, one can essentially preorder the number and masses of solitons or of black holes with scalar hair in these theories, simply by choosing the appropriate boundary condition function $W$. Designer gravity 
theories also have interesting cosmological applications, because certain $W$ admit solutions where smooth 
asymptotically AdS initial data evolve to a big crunch singularity \cite{Hertog03c,Hertog04b}. In supergravity theories 
with a dual conformal field theory (CFT) description one can study the quantum nature of this singularity\footnote{See \cite{Hertog05b,Elitzur06} for recent work on this.} using the AdS/CFT correspondence \cite{Aharony00}. 

The AdS/CFT duality relates $W$ to a potential term $\int W({ O})$ in the dual CFT action, where ${\cal O}$ is the field 
theory operator that is dual to the bulk scalar \cite{Witten02, Berkooz02}. This led \cite{Hertog05} to conjecture that (a) there is a lower bound on the gravitational energy in those designer gravity theories where $W$ is bounded from below, and that (b) the solutions locally minimizing the energy are given by the spherically symmetric, static soliton configurations found in 
\cite{Hertog05}.

Following these conjectures, the stability of designer gravity theories has been studied using purely gravitational 
arguments. A lower bound on the conserved energy in terms of the global minimum of $W$ was established in \cite{Hertog05c} for a consistent truncation of ${\cal N}=8$ $D=4$ gauged supergravity that has several $m^2=-2$ scalars. 
This bound was obtained by relating the Hamiltonian charges to the spinor 
charges, which were shown to be positive for all $W$. It was further conjectured in \cite{Hertog05c} that this result should generalize to all designer gravity theories in $d$ dimensions where the scalar potential $V$ arises from a superpotential $P$, and the scalar reaches an extremum of $P$ at infinity. A more detailed derivation that seemed to confirm this was subsequently given in \cite{Amsel06}.

In this paper, however, we present negative mass solutions in theories where $V$ can be written in terms of a 
superpotential $P$ for boundary conditions specified by a positive function $W$. These solutions are constructed by 
conformally rescaling spherical static solitons that obey designer gravity boundary conditions which preserve the full 
AdS symmetry group. Since the energy can be made arbitrary small, these findings suggest that only a subclass of 
superpotentials have a stable (purely bosonic) ground state when $W$ is bounded from below.

%
\section{Designer Gravity}
%

\subsection{Tachyonic Scalars in AdS}

We consider gravity in $d \geq 4$ spacetime dimensions minimally coupled to a scalar 
field with potential $V(\phi)$. So the action is
\be \label{act}
S=\int d^d x \sqrt{-g} \left [{1\over 2} R - {1\over 2}(\nabla\phi)^2 -V(\phi)\right ]
\ee
where we have set $8\pi G=1$. We require the potential can be written as
\be \label{superpot}
V(\phi) = (d-2) P'^2 - (d-1)P^2
\ee
for some function $P(\phi)$. Scalar potentials of this form arise in the context of $N=1$ supergravity 
coupled to $N=1$ matter, in which case $P(\phi)$ is the superpotential. We are interested in configurations 
where $\phi$ asymptotically approaches a positive local minimum of $P$ at $\phi=\phi_0$. An extremum 
of $P$ is always an extremum of $V$, and (\ref{superpot}) implies that $V_0=V(\phi_0) < 0$. 
Hence $\phi=\phi_0$ corresponds to an anti-de Sitter solution, with metric
\be \label{adsmetric}
ds^2_0 = \bar g_{\mu \nu} dx^{\mu} dx^{\nu}=
-(1+{ r^2 \over l^2})dt^2 + {dr^2\over 1+{ r^2 \over l^2}} + r^2 d\Omega_{d-2}
\ee
where the AdS radius is given by
\be
l^2 = -{(d-1)(d-2) \over 2 V_0}
\ee
At an extremum of $P$ one has
\be\label{ddsuperpot}
V'' = 2P'' \left[ (d-2)P'' -(d-1)P\right]
\ee
so a positive local minimum of $P$ corresponds to a minimum of $V$ only when $(d-2)P_0'' > (d-1)P_0$.
This is a quadratic equation for $P''$, which has a real solution if only if
\be
V'' \geq - {(d-1)^2 P^2 \over 2(d-2)} = {(d-1)V_0 \over 2(d-2)} = - {(d-1)^2 \over 4l^2}
\ee
Hence we recover the BF bound $m^2_{BF}= - {(d-1)^2 \over 4l^2}$ on the scalar mass, which is required for the AdS 
solution to be perturbatively stable. 

We will focus on positive extrema of superpotentials where
\be \label{Wrange}
(d-1)P_0< 2(d-2)P_0'' < (d+1)P_0
\ee
These correspond to tachyonic scalars in AdS with mass $m^2$ in the range
\be \label{range}
m^2_{BF} < m^2 < m^2_{BF}+{ 1 \over l^2}<0.
\ee
Solutions to the linearized wave equation $\nabla^2 \phi - m^2 \phi=0$ for tachyonic scalars in an AdS background, 
with harmonic time dependence $e^{-i\omega t}$, all fall off asymptotically like\footnote{For fields that saturate the 
BF bound, $\l_{+}=\l_{-} \equiv \l$ and $\phi = {\alpha \over r^{\lambda}}\ln r  + {\beta \over r^{\lambda}}$.}
\be\label{genfall}
\phi -\phi_0= {\alpha \over r^{\lambda_{-}}}  + {\beta \over r^{\lambda_{+}}}
\ee
where $\a$ and $\b$ are functions of $t$ and the angles and
\be\label{fallofftest}
\lambda_\pm = {d-1 \over 2} \pm {1 \over 2} \sqrt{(d-1)^2 + 4l^2 m^2}.
\ee
When the scalar mass lies in the range (\ref{range}) both modes are normalizable and hence
should a priori represent physically acceptable fluctuations. To have a well defined theory, however,
one must specify boundary conditions at spacelike infinity.
In general this amounts to choosing a functional relation between $\a$ and $\b$.
The standard choice of boundary condition corresponds to taking $\a=0$ in (\ref{genfall}). 
Taking in account the self-interaction of the scalar field, as well as its backreaction on the geometry, 
one finds this is consistent with the usual set of asymptotic conditions 
on the metric components that is left invariant under $SO(d-1,2)$ \cite{Henneaux85}. In particular, writing 
the metric as $g_{\mu \nu} = \bar g_{\mu \nu} +h_{\mu \nu}$, the asymptotic behavior of the gravitational 
fields is given by
\be \label{asmetric}
h_{rr}={\cal O}(r^{-(d+1)}), \qquad h_{rm}={\cal O}(r^{-d}), \qquad h_{mn}={\cal O}(r^{-d+3})
\ee
where the indices $m,n$ label the time coordinate $t$ and the $d-2$ angles.
Furthermore, the charges that generate the asymptotic symmetries $\xi^{\mu}$ 
involve only the metric and its derivatives\footnote{There is a finite contribution to the conserved charges from the scalar field 
if this saturates the BF bound \cite{Hertog03c}.}. They are given by \cite{Henneaux85}
\be\label{gravch}
{\cal H}_{G} [\xi] = \frac{1}{2}\oint d^{d-2} S_i
\bar G^{ijkl}(\xi^\perp \bar{D}_j h_{kl}-h_{kl}\bar{D}_j\xi^\perp)
\ee
where $G^{ijkl}={1 \over 2} g^{1/2} (g^{ik}g^{jl}+g^{il}g^{jk}-2g^{ij}g^{kl})$,
$h_{ij}=g_{ij}-\bar{g}_{ij}$ is the deviation from the spatial $AdS$ metric $\bar{g}_{ij}$,  
$\bar{D}_i$ denotes covariant differentiation with respect to $\bar{g}_{ij}$ and 
$\xi^\perp = \xi \cdot n$ with $n$ the unit normal to the surface.

One can, however, choose different scalar boundary conditions, defined by $\a \neq 0$ and $\b = \b(\a)$
\cite{Henneaux02,Henneaux04,Hertog04,Hertog05,Henneaux06,Amsel06}. Such asymptotic conditions in general break 
the 
AdS symmetry to $\Re \times SO(d-1)$ - the asymptotic scalar profile changes under the action of $\xi^r$ - but the 
conserved charges associated with the remaining asymptotic symmetries are well defined and finite. 
The dynamical properties of designer gravity theories, however, depend significantly on $\b(\a)$. 
Here we are primarily concerned with the positivity properties of the conserved energy, or more generally with the 
conditions that the superpotential $P$ and the boundary condition function $\b(\a)$ must satisfy for the theory to have a stable 
ground state. But first we briefly review the asymptotics and the construction of conserved charges in designer gravity.

\subsection{Asymptotics and Conserved Charges}

The backreaction of the $\a$-branch of the scalar field, as well as its self-interaction, causes the metric 
components $h_{rm}$ to fall off slower than usual. A complete analysis of the asymptotics is given in 
\cite{Henneaux06}, where it is shown that the asymptotic behavior of the scalar and the gravitational fields 
in designer gravity depends not only on the scalar mass $m^2$, but in general also on the cubic, quartic and 
even quintic terms in the scalar potential. It is also found that the asymptotic fields generally develop 
logarithmic branches for `resonant' scalars\footnote{See \cite{Banados06} for recent work in the context of the AdS/CFT 
correspondencen on resonant scalars with $\a \neq 0$ boundary conditions.}, 
i.e. at integer values of the ratio $\lambda_{+}/\lambda_{-}$.

For our purposes, however, it will be sufficient to restrict attention to even scalar potentials of the form
\be \label{potgen}
V(\phi) = \Lambda + {1 \over 2} m^2 \phi^2 +{1 \over 4} C\phi^4 +{\cal  O}(\phi^6) + ...
\ee
with $m^2$ in the range (\ref{range}) and $\Lambda= -{(d-1)(d-2) \over 2}$ so that $l^2 =1$. $C$ is taken to 
be a free parameter for $m^2 < { (d-1)^2 \over 16 l^2}$, but $C={(d-1) \lambda_{-}^2 \over 8 (d-2)} $ otherwise.
For these scalar potentials the analysis of the asymptotics given in \cite{Hertog04} applies. 
In particular, the asymptotic scalar profile, when one takes in account its backreaction on the geometry, 
is given by (\ref{genfall}), with $\b(\a)$. The corresponding asymptotic behavior of the metric 
components that allows the construction of well defined and finite Hamiltonian generators is given by
\be \label{4-grr}
h_{rr}= -\frac{\alpha^2 \lambda_{-}}{(d-2)} l^2r^{-2-2\lambda_{-}}+ {\cal O} (r^{-(d+1)}), \quad 
h_{rm} = {\cal O}(r^{-d+2}) 
\ee
The expression for the conserved charges depends on the asymptotic behavior of the fields, and is defined 
as follows. Let $\xi^{\mu}$ be an asymptotic Killing vector field. The Hamiltonian takes the form
\be
{\cal H}[\xi] = \int_{\Sigma}^{\ } \xi^{\mu} C_{\mu} + {\rm surface} \ {\rm terms}
\ee
where $\Sigma$ is a spacelike surface, $C_{\mu}$ are the usual constraints, and the surface terms should be 
chosen so that the variation of the Hamiltonian is well defined. The variation of the usual gravitational 
surface term (\ref{gravch}) diverges as $\sim r^{d-(2\lambda_{-}+1)}$ if $\a \neq 0$, 
but there is an additional contribution to the variation of the surface terms that involves the scalar field
\be\label{scalch}
\delta {\cal H}_\phi[\xi] =-\oint \xi^\perp \delta \phi D_i \phi dS^i 
\ee
which exactly cancels the divergence of the gravitational term \cite{Hertog04,Hertog05,Henneaux06,Amsel06}.
The total charge can therefore be integrated, which yields
\be\label{ch}
{\cal H} [\xi] = {\cal H}_{G}[\xi]+ { \lambda_{-} \over (d-2)} r^{d-(2\lambda_{-}+1)} \oint \a^2 d\Omega  
+(\lambda_{+} - \lambda_{-})\oint \left[ {\lambda_{-} \over  (\lambda_{+} - \lambda_{-})} \a\b 
+ W(\a) \right]d\Omega 
\ee
where we have defined a smooth function
\be \label{bc}
W(\a) = \int_0^\a \b(\tilde \a) d \tilde \a 
\ee
which specifies the choice of boundary conditions in designer gravity.
For spherically symmetric solutions, therefore, a manifestly finite expression for the mass ${\cal H} [\partial_t] $ is given by
\be\label{mass}
M = {\rm Vol}(S^{d-2}) \left[{(d-2) \over 2} M_0 + \lambda_{-} \a\b +(\lambda_{+} - \lambda_{-}) W \right]
\ee
where $M_0$ is the coefficient of the $1/r^{d+1}$ term in the asymptotic expansion (\ref{4-grr})
of $h_{rr}$.

\subsection{Stability and Ground State}

By generalizing Witten's spinorial proof of the Positive Energy Theorem (PET) for asymptotically flat 
spacetimes \cite{Witten81}, it has been shown \cite{Townsend84} (see also \cite{Gibbons83}) that for 
the case of a single scalar field and with standard $\a=0$ 
scalar boundary conditions, a potential $V(\phi)$ admits a PET {\it if and only if} $V$
can be written in terms of a 'superpotential' $P(\phi)$, and $\phi$ approaches an extremum of $P$ at infinity. 

This result obviously concerns the positivity properties of the energy as defined by the spinor charge 
\be \label{spinorcharge}
{\mathcal Q}_\xi = \oint *{\bf B}
\ee 
where the integrand is the dual of the Nester 2-form, with components
\be\label{2form}
{B}_{ab} = 
\frac{1}{2}(\overline \psi \gamma^{[c} \gamma^d \gamma^{e]} \widehat \nabla_e \psi 
+ {\rm h.c.})\epsilon_{abcd} \, 
\ee
$\psi$ is taken to be an asymptotically supercovariantly constant spinor field and
\be\label{covder}
\widehat \nabla_a \psi 
= \left[ \nabla_a - {1 \over \sqrt{2(d-2)}} \gamma_a P(\phi) \right]\psi
\ee
with $P(\phi)$ given by (\ref{superpot}). This definition of the covariant derivative enabled \cite{Townsend84} to 
express the spinor charge (\ref{spinorcharge}) as a manifestly non-negative quantity, provided $\psi$ 
satisfies the spatial Dirac equation $\gamma^i \widehat D_i \, \psi = 0$. In the context of $N=1$ supergravity $P$ is the superpotential, but the argument of \cite{Townsend84} applies to any gravity plus scalar theory, irrespective of whether it is 
a sector of a supergravity theory. 

Townsend's Positive Energy Theorem \cite{Townsend84} establishes the positivity of the Hamiltonian generator 
(\ref{gravch}), because this equals the spinor charge for $\a=0$. This is not the case in designer gravity, however,
where the $\a$ branch of the scalar modifies the expression of the charges. Indeed it follows from the asymptotic expansion of the spinor field, and the asymptotic expansions of the metric and the scalar field, that the Hamiltonian charges (\ref{ch}) are related to the spinor charges (\ref{spinorcharge}) as \cite{Hertog05c}
\be
\label{hq}
{\cal H}_\xi = {\mathcal Q}_\xi +(\lambda_{+} - \lambda_{-}) \oint W (\alpha) d \Omega
\ee
where $ W(\a)$ is defined in (\ref{bc}).
The spinor charge, therefore, needs not be conserved in designer gravity. Instead it depends on the choice of 
the cross section $S^{d-2}$ at infinity, because $\alpha$ is, in general, time dependent. On the other hand, the above calculation that leads from the Witten condition to the positivity of the spinor charge still applies. Taking
$\xi=\partial_t +\omega \partial_{\phi}$, with $\vert \omega \vert <1$, this yields \cite{Hertog05c,Amsel06}
\be
E + \omega J \ge (\lambda_{+} - \lambda_{-})\oint W(\alpha) d \Omega
\ee
and therefore
\be\label{bound}
 E \ge {\rm Vol}(S^{d-2}) (\lambda_{+} - \lambda_{-})\, {\rm inf} \, W + |J|.
\ee  
where $J$ is the angular momentum. Hence it would seem to follow that the energy is bounded from below in designer gravity theories (for scalar potentials that arise from a superpotential $P$ and with $m^2$ in the range (\ref{range})) for all asymptotic 
conditions (\ref{bc}) that are defined by a function $ W(\a)$ that has a global minimum. Furthermore, the inequality (\ref{bound}) suggests that theories where $ W$ is unbounded from below admit smooth solutions with arbitrary negative energy. Such solutions have indeed been shown to exist in certain theories \cite{Hertog04b}. The only subtlety in the derivation of (\ref{bound}) is showing that with $\a \neq 0$ boundary conditions, asymptotically supercovariantly constant solutions to $\gamma^i \widehat D_i \, \psi = 0$ exist. This was shown for the consistent truncation of  ${\cal N}=8$ $d=4$ gauged supergravity studied in \cite{Hertog05c}, but this has not been demonstrated in general. We return to this point in the conclusion.

Finally we mention that in the case of supergravity theories with a dual field theory description, the AdS/CFT correspondence indicates \cite{Hertog05} that the true ground state of the theory (when $ W$ is bounded from below) is given by the lowest energy spherical soliton. The nature of the ground state has not been established yet, however, using purely gravitational arguments\footnote{The lowest energy soliton does not saturate the lower bound (\ref{bound}), because the actual soliton mass has an additional positive contribution coming from the spinor charge.}.

\subsection{AdS-invariant boundary conditions}

Designer gravity boundary conditions generally break the asymptotic AdS symmetry to $\Re \times SO(d-1)$. 
The full AdS symmetry group is preserved, however, for asymptotic conditions defined by
\be
 W(\a)=k \a^{d-1/\lambda_{-}}
\ee
where $k$ is an arbitrary constant without variation\footnote{We note that this defines AdS-invariant boundary conditions
only for scalar potentials of the form (\ref{potgen}). The expression of conformally invariant asymptotics for
other potentials is given in \cite{Henneaux06,Amsel06}.}. In this case the total charge (\ref{ch}) becomes
\be\label{adsch}
{\cal H} [\xi] = {\cal H}_{G}[\xi]+ { \lambda_{-} \over (d-2)} r^{d-(2\lambda_{-}+1)} \oint \a^2 d\Omega  
+2k\lambda_{+} \oint \a^{d-1/\lambda_{-}}  d\Omega 
\ee
which yields the following expression for the mass of spherically symmetric solutions,
\be\label{adsmass}
M = {\rm Vol}(S^{d-2}) \left[{(d-2) \over 2} M_0 + 2 k \lambda_{+}  \a^{d-1/\lambda_{-}} \right]
\ee
In the next section we use the conformal rescaling symmetry of this class of boundary conditions to 
show there are superpotentials for which the energy bounds (\ref{bound}) do not hold.

%
\section{Violation of Energy Bounds}
%

\subsection{Asymptotically AdS Solitons}

Consider the following class of superpotentials in $d=4$ dimensions,
\be\label{superpot2}
P(\phi)= (1+{1 \over 2} \phi^2 ) e^{-{A \over 4} \phi^4}
\ee
where $A >0 $ is a free parameter. These yield scalar potentials with a negative maximum at $\phi=0$, and with two global minima at $\phi = \pm \phi_m$. The potential corresponding to (\ref{superpot2}) with $A=1/4$ is plotted in Figure 1. Small fluctuations around $\phi=0$ have $m^2 =-2$, which is above the BF bound and within the range (\ref{range}). Hence asymptotically the scalar generically decays as
\be\label{genfall2}
\phi = {\alpha \over r }  + {\beta \over r^2}
\ee
and the asymptotic behavior of the $g_{rr}$ metric component reads
\be\label{asmetric2}
g_{rr} = {1 \over r^2} - { (1+\a^2/2 )\over r^4} +{\cal O} (r^{-5})
\ee

\begin{figure}
\begin{picture}(0,0)
\put(225,200){$V$}
\put(390,136){$\phi$}
\end{picture}
\centerline{\epsfig{file=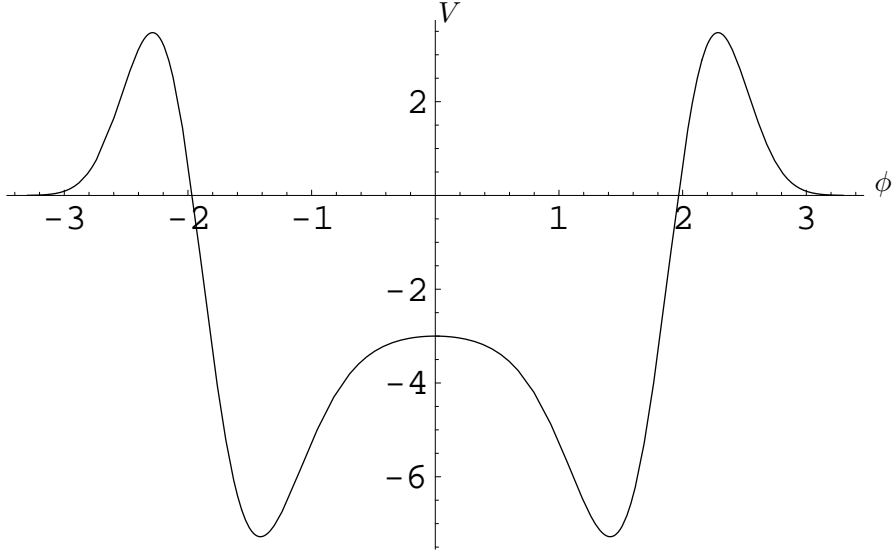,width=4.5in}}
\caption{Scalar potential $V$ that can be written in terms of a superpotential $P$ with $P'(0)=0$.}
\label{1}
\end{figure}

We adopt AdS-invariant boundary conditions defined by $ W(\a)=0$ everywhere. The conserved mass, therefore, is simply given by the surface integral of the coefficient of the $1/r^5$ term in (\ref{asmetric2}). According to the lower bound (\ref{bound}) this should be positive for all solutions where $\phi$ asymptotically decays as $\phi \sim \a/r +{\cal O}(1/r^3)$. 
We now show, however, that for a wide range of values of $A$ there are negative mass solutions.

We begin by looking for static spherical soliton solutions of the theory (\ref{superpot2}). Writing the metric as
\be
ds^2=-h(r)e^{-2\chi(r)}dt^2+h^{-1}(r)dr^2+r^2d\Omega_2
\ee
the field equations read
\be\label{hairy14d}
h\phi_{,rr}+\left(\frac{2h}{r}+\frac{r}{2}\phi_{,r}^2h+h_{,r} \right)\phi_{,r}   =  V_{,\phi}
\ee
\be\label{hairy24d}
1-h-rh_{,r}-\frac{r^2}{2}\phi_{,r}^2h =  r^2V(\phi)
\ee
\be \label{hairy34d}
\chi_{,r} = -{1 \over 2}r \phi_{,r}^2
\ee
Regularity at the origin requires $h=1$ and $h_{,r}=\phi_{,r}=\chi_{,r}=0$ at $r=0$.
Rescaling $t$ shifts $\chi$ by a constant, so its value at the origin is arbitrary. 
Thus solutions can be labeled by the value of $\phi$ at the origin.

\begin{figure}
\begin{picture}(0,0)
\put(36,205){$\phi$}
\put(335,20){$r$}
\end{picture}
\centering{\psfig{file=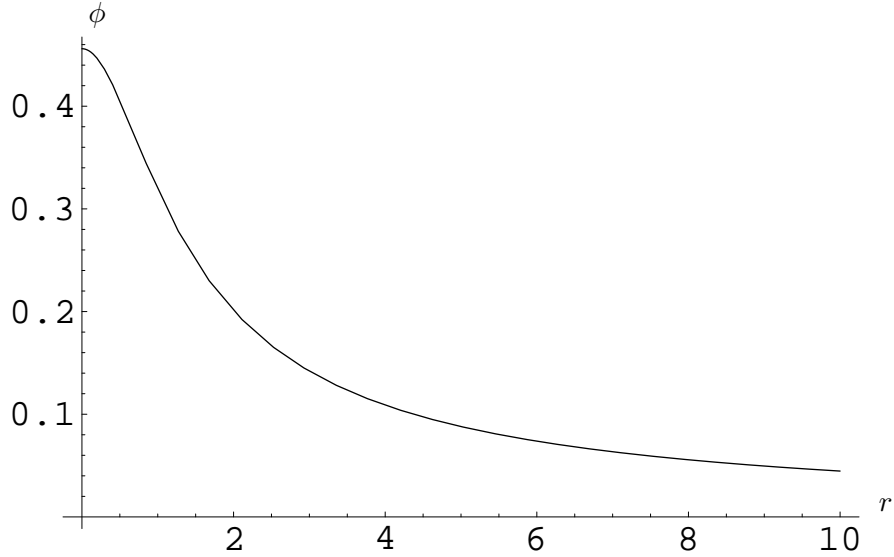,width=4.5in}}
\caption{Soliton solution $\phi (r)$ with boundary conditions specified by $\b=0$.}
\label{2}
\end{figure}

One can numerically integrate the field equations. For every nonzero $\phi(0)$ at the origin in the range 
$-\phi_m < \phi (0) < \phi_m$, the solution to (\ref{hairy14d}) is asymptotically of the form (\ref{genfall2}).
The staticity and spherical symmetry of the soliton mean $\a(t,\Omega)$ and $\b(t,\Omega)$ are simply 
constants. For $A \sim {\cal O}(1)$ we find there is a `critical' value $\phi_c (0)$ for which $\b=0$, and hence 
$\phi \sim \a/r +{\cal O}(1/r^3)$ asymptotically. We plot this soliton solution $\phi_{s}(r)$ 
in Figure 2 for the $A=1/4$ potential. We have found a class of scalar potentials, therefore, that can be derived from a 
superpotential and admit regular static spherical soliton solutions for AdS-invariant boundary conditions.

\subsection{AdS solitons imply negative energy}

The existence of scalar solitons with AdS-invariant boundary conditions implies there are negative mass 
solutions in these theories. This was shown, using scaling arguments, in \cite{Heusler92} for
non-negative potentials and then generalized to potentials with a negative local maximum in
\cite{Hertog04b}. We emphasize the claim is not that the soliton itself must have negative energy (in general 
it has positive mass), but only that negative energy solutions must exist. 

To apply the scaling arguments of \cite{Heusler92,Hertog04b} to our case we first need an explicit formula 
for the mass of spherically symmetric (and time symmetric) initial data when the scalar field has a profile 
$\phi(r)$. In this case, the constraint equations reduce to 
\be\label{constr}
\ ^{3}{\cal R} =  g^{ij}\phi_{,i}\phi_{,j} + 2 V(\phi) 
\ee
Writing the spatial metric as
\be \label{metric}
ds^2 = \left(1-{m(r)\over r}+r^2 \right)^{-1} dr^2 + r^2 d\Omega_{2}
\ee
the constraint (\ref{constr}) yields the following equation for $m(r)$ 
\be \label{mscalar}
m_{,r} +\frac{1}{2}m(r)r\phi_{,r}^2 = r^2
\left[(V(\phi)-\Lambda)+{1 \over 2} \left(1+ r^2\right)\phi_{,r}^2 \right]
\ee
The general solution for arbitrary $\phi (r)$ is
\be\label{gensoln} 
m(r) = \int_{0}^{r} e^{-{1\over 2}\int_{\tilde r}^r  d\hat r \ \hat r\phi_{,\hat r}^2} 
\left[(V(\phi)-\Lambda) +{1 \over 2} \left(1+ \tilde r^2 \right)
\phi_{,\tilde r}^2 \right] \tilde r^{2} d\tilde r.  
\ee
Hence the total mass (\ref{ch}) is given by
\be \label{totm}
M = 4\pi \lim_{r\to\infty} \left[ m(r) +{ \a^2 \over 2} r\right]
\ee

Now suppose $\phi_s(r)$ is a static soliton and consider the one parameter family of configurations 
$\phi_\l(r) = \phi_s(\l r)$. Because of the conformal rescaling symmetry these obey the same boundary 
conditions as the soliton. Then from (\ref{gensoln}) and (\ref{totm}), it is easy to see that the total 
mass of the rescaled configurations takes the form
\be
M_\l = \l^{-3} M_1 + \l^{-1} M_2
\ee
where $M_2$ is independent of the potential and is manifestly positive, and
both $M_i$ are finite and independent of $\l$. 
Furthermore, because the static soliton extremizes the energy \cite{Sudarsky92} one has
\be
0={d M_\l\over d\l}|_{\l=1} = -3 M_1 -M_2
\ee
and hence $M_1=-{1 \over 3} M_2 <0$. 
Therefore the contribution to the mass that scales as the volume, which includes the potential and scalar 
terms, is negative. This means that rescaled configurations $\phi_\l(r)$ with $\l < 1/\sqrt{3}$ must 
have negative total mass\footnote{We have verified that the rescaled configurations $\phi_\l (r)$ are regular initial data
for small $\l$, i.e. $h_\l(r)= r^2 +1 - {m_\l(r) \over r}$ is strictly positive everywhere.}, 
and hence violate the energy bound 
(\ref{bound}). For the soliton solution shown in Figure 2 we find $M_1=-{1 \over 3} M_2=-1/4$, and hence $M=1/2$. 

The rescaled configurations are initial data for time dependent solutions. 
For sufficiently small $\l$ one has a large central region where $\phi$ is essentially constant and away 
from an extremum of the potential. Hence one expects the field to evolve to a spacelike singularity. 
This singularity cannot be hidden behind an event horizon, because the mass of all spherically 
symmetric black holes is larger than the soliton mass. Instead, one expects initial data of this type to 
produce a big crunch\footnote{Initial data for which this can be shown rigorously can be constructed from Euclidean 
$O(4)$-invariant instanton solutions of the form $ds^2 = { d\rho^2 \over b^2 (\rho)} +\rho^2 d\Omega_3$. The slice through 
the instanton obtained by restricting to the equator of the $S^3$ defines time symmetric initial data for a zero mass
Lorentzian solution. With conformally invariant boundary conditions, the evolution of these initial data is simply obtained 
from analytic continuation of the instanton geometry. One finds the spacetime evolves like a collapsing FRW universe
\cite{Hertog04b,Hertog05b}. For the $A=1/4$ potential of the form (\ref{superpot2}) we have considered here, the instanton 
that obeys $W=0$ boundary conditions has $\phi(0)=.652$.} \cite{Hertog04b,Hertog05b}.

\subsection{Further Examples}

Finally we show that $W=0$ is not an isolated example of boundary conditions for which the bounds (\ref{bound})
do not hold. Consider AdS gravity coupled to a scalar with $m^2 =-27/16$ in four dimensions, with AdS-invariant 
boundary conditions defined by
\be \label{genadsbc}
 W(\a)={ k \over 4}  \a^4
\ee
where $k$ is an arbitrary constant. According to (\ref{bound}) the theory should satisfy the PET
when $k \geq 0$. We find below, however, that negative mass solutions exist for all $k$. 

We concentrate on the following class of potentials,
\be\label{pot3}
V(\phi)=-3-{27 \over 32} \phi^2 - {27 \over 256} \phi^4 -{3 \over 4} \phi^6 +B \phi^8
\ee
where $B$ is a free parameter. For positive $B$ these are qualitatively similar to the potentials 
we considered above, with a negative maximum at $\phi=0$ and global minima at $\phi = \pm \phi_m$. But
scalar fluctuations around $\phi=0$ now have mass $m^2 =-27/16$, so the scalar generically decays as
\be\label{genfall3}
\phi = {\alpha \over r^{3/4} }  + k{\a^3 \over r^{9/4}}.
\ee
Townsend's result \cite{Townsend84} says that potentials of this form admit the PET for solutions that 
asymptotically behave as $\phi \sim 1/r^{9/4}$, if (and only if) $V$ can be derived from a superpotential 
$P$ with $P'(0)=0$. To construct the corresponding superpotential one needs to solve
\be\label{Weq}
P'(\phi) =\frac{1}{\sqrt{2}}\sqrt{V+3 P^2} 
\ee
starting with $P(0)=1$.

\begin{figure}
\begin{picture}(0,0)
\put(225,200){V}
\put(392,159){$\phi$}
\end{picture}
\centerline{\epsfig{file=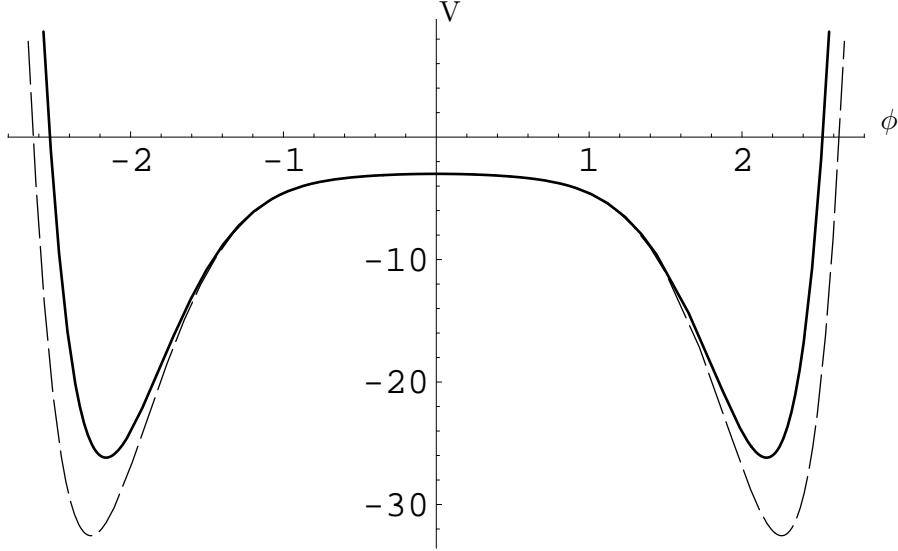,width=4.5in}}
\caption{The dashed line corresponds to a critical scalar potential $V$ that is on the verge of violating 
the Positive Energy Theorem for standard scalar AdS boundary conditions. The full line gives a potential that 
arises from a superpotential, yet violates the PET with $W>0$ designer gravity boundary conditions}
\label{3}
\end{figure}

A solution to (\ref{Weq}) exists unless the quantity inside the square root becomes negative. As we integrate 
out from $\phi=0$, $P$ is increasing and the square root remains real because the scalar satisfies the BF bound. 
For sufficiently large values of $B$ the global minima at $\pm \phi_m$ will not be very much lower than the local 
maximum at $\phi=0$, so a global solution for $P$ will exist and $P'(\phi_m)>0$. This is expected, 
since the PET (for $\a=0$) holds for potentials of this form. 
If the global minima are too deep, however, the quantity under the square root will become negative before 
the global minimum is reached, and a real solution will not exist. Clearly the critical potential corresponds 
to one where $V+3 P^2$ just vanishes as the global minimum is reached. In other words, the condition for a 
potential $V$ to be on the verge of violating the PET is simply $P'(\phi_m)=0$. 
We find that the critical potential of the form (\ref{pot3}) has $B_c=.1138$. For $B < B_c$ the PET does 
not hold for solutions where $\phi \rightarrow 0$ at infinity, whereas scalar potentials (\ref{pot3}) with 
$B \geq B_c$ can be written in terms of a superpotential, and hence admit the PET for solutions where 
$\phi \sim 1/r^{9/4}$ asymptotically. We plot the critical potential in Figure 3 
(dashed curve), as well as the $B=.125$ potential whose properties we discuss in more detail below.

In the regime where a superpotential exists, the lower bounds (\ref{bound}) would imply that the 
theory should satisfy the PET not only for $\a=0$, but also for generalized 
AdS-invariant boundary conditions (\ref{genadsbc}) with $k \geq 0$. We now show, however, there are 
$B > B_c$ for which (\ref{pot3}) admits exactly one regular static spherical soliton solution for 
all $k \geq 0$. This means the bounds (\ref{bound}) cannot hold, because one can again 
conformally rescale the asymptotically AdS solitons to construct negative mass initial data.

The set of soliton solutions of a particular potential with a negative maximum is found by integrating the 
field equations (\ref{hairy14d})-(\ref{hairy34d}) for different values of $\phi$ at the origin. 
For $\phi(0)$ in the range $-\phi_m < \phi (0) < \phi_m$ the scalar asymptotically behaves as (\ref{genfall3}) 
so we get a point in the $(\a,\b)$ plane. Repeating for all $\phi (0)$ yields a curve $\b_{s}(\a)$. Given a 
choice of boundary condition $\b(\a)$, the allowed solitons are simply given by the points where the soliton 
curve intersects the boundary condition curve: $\b_{s}(\a)=\b(\a)$.

\begin{figure}
\begin{picture}(0,0)
\put(90,201){$\b$}
\put(388,22){$\a$}
\end{picture}
\centerline{\epsfig{file=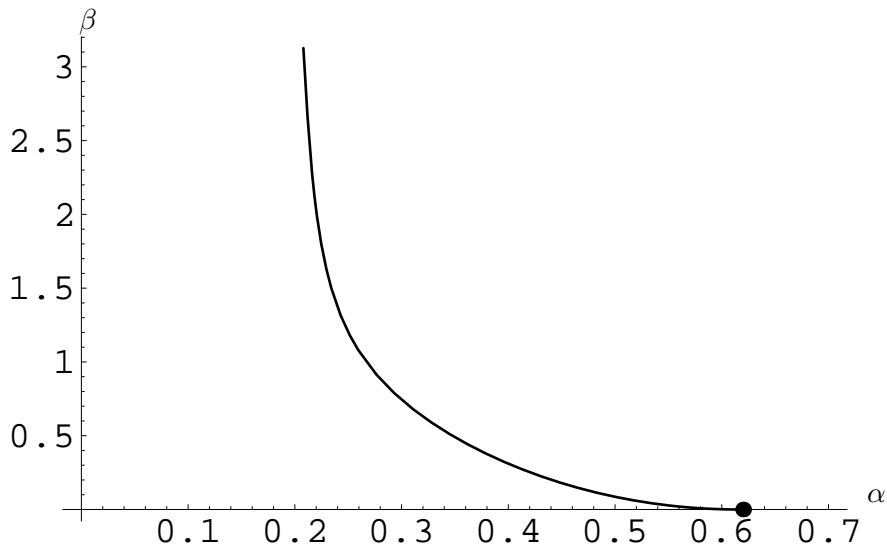,width=4.5in}}
\caption{The function $\b_{s}(\a)$ obtained from the solitons.}
\label{4}
\end{figure}

A section of the soliton curve $\b_{s}(\a)$ for the $B=.125$ potential is plotted in Figure 4.
Along the curve $\phi (0)$ increases from $\phi(0) \approx 1.32$, which corresponds to $\b_{s}=0$, to the global 
minimum at $\phi_m=2.16$ where $\b_{s} \rightarrow \infty$. One sees that for all $k \geq 0$ the soliton curve 
$\b_{s}(\a)$ has precisely one intersection point with the boundary condition function $\b=k\a^3$. 
Hence the conformally rescaled configurations $\phi_{\l}( r)=\phi_{s}(\l r)$ provide, for $\l < 1/\sqrt{3}$, 
examples of negative mass initial data.

When $\phi \rightarrow \phi_m$ one has $\a \rightarrow .2$ in the $B=.125$ theory. This limiting value of $\a$ 
decreases towards zero, however, for $B \rightarrow B_c$. Furthermore, when $B < B_c$ the soliton curve intersects the 
$\a=0$ axis at finite 
$\b$, yielding a regular asymptotically AdS soliton solution for standard $\a=0$ boundary conditions.
This is not surprising, because the potential cannot be derived from a superpotential when $B <B_c$, and hence the 
PET cannot hold \cite{Townsend84}.

%
\section{Conclusion}
%

We have studied the stability of designer gravity theories, where one considers AdS gravity coupled to a scalar 
field with mass at or slightly above the BF bound and with boundary conditions specified by an essentially arbitrary 
function $W$.

By conformally rescaling spherical static solitons that obey AdS-invariant boundary conditions specified by a 
non-negative function $W$, we have constructed solutions with arbitrary negative mass in a class of theories 
where the scalar potential $V$ arises from a superpotential $P$, and $\phi$ reaches an extremum of $P$ at infinity. 
These solutions violate the lower bounds (\ref{bound}) on the conserved energy that were obtained in \cite{Amsel06}, and 
they indicate that this class of theories does not have a stable ground state. We expect that similar instabilities can 
be found in designer gravity theories in $d>4$ dimensions, and for boundary conditions $W$ that break the asymptotic 
AdS symmetry to $\Re \times SO(d-1)$.

The derivation of the lower bounds (\ref{bound}) relies crucially on the positivity of the spinor charge 
in designer gravity. Our findings suggest, therefore, that superpotentials for which these bounds do not hold, do not 
admit asymptotically supercovariantly constant spinor solutions to the spatial Dirac equation, at least for some 
designer gravity boundary conditions. This argument has been advanced long ago in \cite{Hawking83}.
It would be interesting to clarify this point, and to identify the precise criteria that $P$ and $W$ 
must satisfy in order for these spinor solutions to exist.

In this context we should mention that we have found no examples of supergravity theories that violate the energy bounds
and that have a dual description in terms of a field theory which is supersymmetric for $W=0$. Hence the 
positive energy conjectures of \cite{Hertog05} appear to be correct when restricted to this class of theories 
with an AdS/CFT dual. In fact, the 
lower bounds (\ref{bound}) seem rather natural from the point of view of the dual field theory. Remember that 
imposing $ W \neq 0$ boundary conditions on one (or several) tachyonic bulk scalars corresponds to adding a 
potential term $\int W({\cal O})$ to the dual CFT action, where ${\cal O}$ is the field theory operator that is 
dual to the bulk scalar \cite{Witten02, Berkooz02}.
The change in the energy by this deformation is $\oint   < W(O(x))>  d \Omega$, which leads in the 
large $N$ limit - which corresponds to the supergravity approximation - to $\oint  W(<O>)  d \Omega$. 
This clearly leads to (\ref{bound}) provided all configurations in the dual CFT with $ W=0$ satisfy $E \geq |J|$. 

The AdS/CFT correspondence even suggests one should be able to generalize the bounds (\ref{bound}) to
certain classes of $W$ that are unbounded from below. Indeed, the precise correspondence between solitons and field theory 
vacua is captured by the following function \cite{Hertog05},
\be \label{effpot}
{\cal V}(\alpha) = -\int_{0}^{\a} \b_{s} (\tilde \a) d\tilde \a + W(\alpha)
\ee
where $\b_{s}(\a)$ is the function obtained from the set of soliton solutions. It can be shown \cite{Hertog05} that for any 
$W$ the location of the extrema of ${\cal V}$ yield the vacuum expectation values $ \langle {\cal O} \rangle = \a$, and that 
the value of ${\cal V}$ at each extremum yields the energy of the corresponding soliton.
This suggests there should be a lower bound on the energy in all designer gravity theories where 
${\cal V}(\a)$ has a global minimum. For this it is sufficient that $\b_{s} <W' $ at large $\a$. This includes a class of 
boundary condition functions $W$ that are unbounded from below, since $\b_{s}(\a) <0$ for $\a >0$ in theories where
(\ref{bound}) holds.

\bibliography{ref}
\bibliographystyle{JHEP}


\end{document}